\documentclass[
 aip,
 amsmath,amssymb,
 reprint,%
]{revtex4-1}

\usepackage{graphicx}% Include figure files
\usepackage{dcolumn}% Align table columns on decimal point
\usepackage{bm}% bold math
\usepackage[utf8]{inputenc}
\usepackage[T1]{fontenc}
\usepackage{mathptmx}
\usepackage{etoolbox}
\usepackage{xcolor}
\usepackage[normalem]{ulem}
\makeatletter
\def\@email#1#2{%
 \endgroup
 \patchcmd{\titleblock@produce}
  {\frontmatter@RRAPformat}
  {\frontmatter@RRAPformat{\produce@RRAP{*#1\href{mailto:#2}{#2}}}\frontmatter@RRAPformat}
  {}{}
}%
\makeatother
\begin{document}

\preprint{AIP/123-QED}

\title{Out-of-plane spin-to-charge conversion at low temperatures in graphene/MoTe$_{2}$ heterostructures}
% Force line breaks with \\
\author{Nerea Ontoso}
 \affiliation{CIC nanoGUNE BRTA, 20018 Donostia-San Sebastian, Basque Country, Spain}%Lines break automatically or can be forced with \\
 
 \author{C.K. Safeer}
 \affiliation{CIC nanoGUNE BRTA, 20018 Donostia-San Sebastian, Basque Country, Spain}
\affiliation{Present address: Department of Physics, Clarendon Laboratory, University of Oxford, Oxford, United Kingdom}
 \author{ Josep Ingla-Aynés}
 \affiliation{CIC nanoGUNE BRTA, 20018 Donostia-San Sebastian, Basque Country, Spain}
 \affiliation{Present address: Kavli Institute of Nanoscience, Delft University of Technology, 2628 CJ Delft, The Netherlands}
 
 \author{Franz Herling}
 \affiliation{CIC nanoGUNE BRTA, 20018 Donostia-San Sebastian, Basque Country, Spain}
 \affiliation{Present address: Catalan Institute of Nanoscience and Nanotechnology (ICN2), 08193 Barcelona, Spain}

 \author{Luis E. Hueso}
 \affiliation{CIC nanoGUNE BRTA, 20018 Donostia-San Sebastian, Basque Country, Spain}
 \affiliation{IKERBASQUE, Basque Foundation for Science, 48009 Bilbao, Basque Country, Spain}
 
\author{M. Reyes Calvo$^*$}%
%\email{reyes.calvo@ua.es}
\affiliation{Departamento de Física Aplicada, Universidad de Alicante, 03690 Alicante, Spain}
\affiliation{Instituto Universitario de Materiales de Alicante (IUMA), Universidad de Alicante, 03690 Alicante, Spain}

\author{F\`elix Casanova$^*$}
 \email[]{Authors to whom correspondence should be addressed: reyes.calvo@ua.es, f.casanova@nanogune.eu}
\affiliation{CIC nanoGUNE BRTA, 20018 Donostia-San Sebastian, Basque Country, Spain}
\affiliation{IKERBASQUE, Basque Foundation for Science, 48009 Bilbao, Basque Country, Spain}
\date{\today}% It is always \today, today,

\begin{abstract}
Multi-directional spin-to-charge conversion - in which spin polarizations with different orientations can be converted into a charge current in the same direction - has been demonstrated in low-symmetry materials and interfaces. This is possible because, in these systems, spin to charge conversion can occur in unconventional configurations in which  %spin polarization and charge current where 
charge current, spin current and polarization do not need to be mutually orthogonal.
%the spin polarization and the spin or charge current directions are not all mutually orthogonal. 
Here, we explore, in the low temperature regime, the spin-to-charge conversion in heterostructures of graphene with the low-symmetry 1T' phase of MoTe$_2$. First, we observe the emergence of charge conversion for out-of-plane spins at temperatures below 100 K. This unconventional component is allowed by the symmetries of both MoTe$_2$ and graphene and likely arises from spin Hall effect in the spin-orbit proximitized graphene. Moreover, we examine the low-temperature evolution of non-local voltage signals arising from the charge conversion of the two in-plane spin polarizations, which have been previously observed at higher temperature. As a result, we report omni-directional spin-to-charge conversion - for all spin polarization orientations - in graphene/MoTe${_2}$ heterostructures at low temperatures.
\end{abstract}

\maketitle

%%%% 

%\begin{quotation}

%\end{quotation}

%\section{\label{sec:level1}First-level heading:\protect\\ The line
%break was forced \lowercase{via} \textbackslash\textbackslash}
Spin-based logic devices offer a promising alternative to the limitations imposed by Joule heating in traditional electronics \cite{zutic2004,manipatruni2019}. These spintronic devices rely on the efficient generation, precise control and detection of spin currents, for which materials hosting spin-to-charge conversion (SCC) phenomena provide a platform. SCC occurs in a variety of systems with high spin-orbit interactions; mediated by the spin Hall effect (SHE) in bulk materials \cite{sinova2015,Valenzuela2006,Kimura2007} or the Rashba-Edelstein (REE) effects \cite{DYakonov1971,DYakonov1971b,edelstein1990}%,rojas2016,kato2004} 
in Rashba interfaces \cite{%edelstein1990,
rojassanchez2013} or at the surface states of topological insulators \cite{mellnik2014,rojas2016}. 

Spin-orbit-mediated SCC effects were initially investigated in zinc-blende semiconductors and pure metals, such as GaAs \cite{kato2004}, Ta \cite{Liu2012} or Pt \cite{sagasta2016}, all systems with a high degree of crystal symmetry. As a consequence, SCC phenomena have been traditionally expected to appear in configurations where charge current, spin current and spin polarization directions are mutually orthogonal. Nonetheless, these restrictions only hold for materials with high crystal symmetry and are lifted for crystalline phases where certain mirror symmetries are broken \cite{culcer2007generation,seemann2015symmetry,zhang2021different,roy2022}. 

Recent advancements in the preparation of atomically thin layers of van der Waals materials have raised the interest in their spintronic properties. In particular, transition metal dichalcogenides (TMDs) exhibit strong spin-orbit interactions and are good candidates for hosting SCC phenomena. Besides mutually orthogonal conventional SCC, non-orthogonal unconventional SCC configurations have been recently reported for TMDs with strong spin-orbit coupling and low-symmetry crystal phases \cite{stiehl2019,SafeerOntoso2019,Ontoso_MoTe2_2022,macneill2017wte2,zhao2020,kao2022}. Furthermore, TMDs have been integrated in van der Waals heterostructures with graphene, which acts as spin channel due to its long spin diffusion length \cite{Tombros2007,InglaAynes2015}. The combination of SHE and REE effects, enabled by the strong spin-orbit proximity from TMDs to graphene, has led to highly efficient conventional SCC in these heterostructures \cite{safeer2019,herling2020,ghiasi2019,benitez2020,li2020,hoque2020,hoque2021,zhao2020c,garcia2018spin,offidani2017optimal}. Additionally, unconventional SCC has been observed in graphene/TMD heterostructures %, even certain components forbidden by the crystal symmetries of both the TMD and the graphene 
\cite{SafeerOntoso2019,camosi_2022_wte2_unc,ingla2022_nbse2_omnidirectional,zhao2020}. Symmetry breaking at the van der Waals interface has been proposed to facilitate unusual non-orthogonal SCC in graphene/TMD heterostructures - even for highly symmetric TMD phases \cite{li2019_twist,david2019,naimer2021twist,peterfalvi2022quantum,veneri2022twist,lee2022charge}. As a result, multi- and omni-directional SCC has been already reported for heterostructures of graphene with MoTe$_2$ \cite{SafeerOntoso2019}, WTe$_2$ \cite{camosi_2022_wte2_unc} and NbSe$_2$ \cite{ingla2022_nbse2_omnidirectional}, enhancing the efficiency of spin generation and detection and increasing the versatility of graphene/TMD spintronic devices. 

Both conventional and unconventional SCC have been observed in the low-crystal-symmetry 1T' phase of MoTe$_2$ and for graphene/MoTe$_2$ \cite{stiehl2019,song2020,hoque2021,SafeerOntoso2019,Ontoso_MoTe2_2022}. In Ref. \cite{SafeerOntoso2019}, we reported SCC in a graphene/MoTe$_2$ heterostructure for the two in-plane spin orientations - one orthogonal and another parallel to the charge current - from 75 K up to room temperature. New SCC processes may emerge at lower temperatures, since conversion in proximitized graphene is enhanced in this regime \cite{ghiasi2019,safeer2019}. In this work, we explore SCC in graphene/MoTe$_2$ lateral spin valve (LSV) devices below 100 K. First, we track the low temperature dependence of both conventional and unconventional SCC for in-plane spin polarization. We find a change of sign of the signal associated to the conventional SCC - allowed by symmetry arguments - that we relate to a change of the relative intensity of two competing effects, namely SHE in MoTe$_2$ and REE in graphene. More interestingly, we detect the SCC of out-of-plane spin polarization, which increases with lowering temperature and likely arises from SHE in the proximitized graphene. Finally, we summarize all the SCC components observed for graphene/MoTe$_2$ in our experimental geometry and discuss on their possible origin. Regardless of the conversion mechanisms, we achieve omnidirectional - for all three spin orientations - SCC at low temperature.

%\textbf{lo he quitado de aqui, pero en algun sitio idea de que out of plane sobrevive hasta 100K y que no lo habiamos visto porque es muy pequena, pero el mejor analisis lo pone en evidencia}

\begin{figure}
\includegraphics[width=\linewidth]{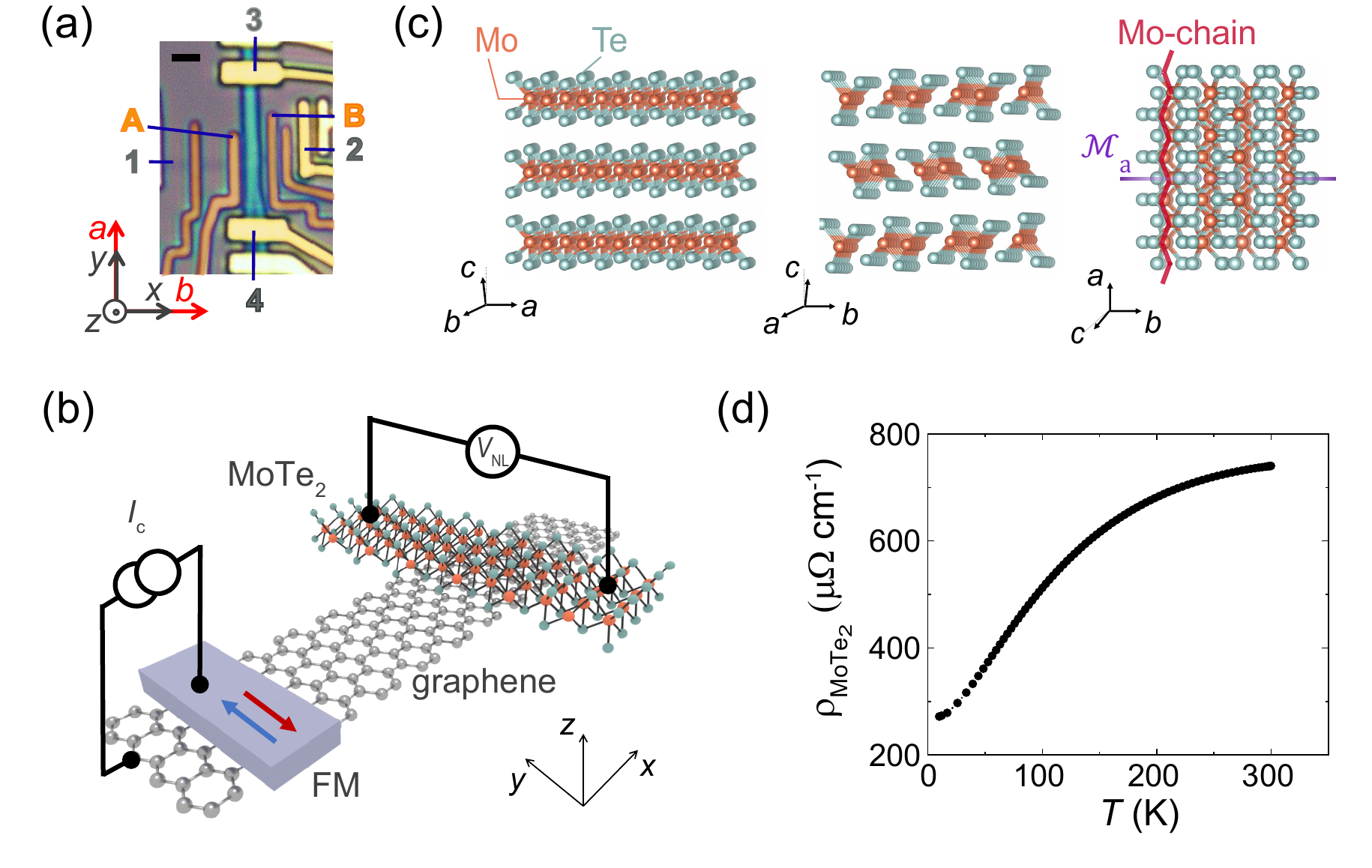}
\caption{\label{fig:1} a) Optical microscope image (scale bar represents 1 $\mu$m) and b) sketch of a graphene/MoTe$_2$ lateral spin valve device. c) Crystal structure of 1T'-MoTe$_2$. Mirror symmetry plane and Mo-chain directions are marked. d) Two-terminal resistance of the MoTe$_2$ flake as a function of temperature measured using leads $3$ and $4$ in a.}
\end{figure}

LSV devices (Fig. 1a,b) were prepared as described in Refs.  \cite{SafeerOntoso2019,Ontoso_MoTe2_2022}. A flake of MoTe$_2$, with thickness ~11 nm, is stacked perpendicularly on top of a graphene channel where magnetic TiO$_x$/Co electrodes (labeled as \textit{A} and \textit{B} in Fig.~\ref{fig:1}a) act as spin injectors (FM in Fig. 1b). Needle-like MoTe$_2$ flakes are typically elongated along the Mo-chain direction (Fig. 1b,c) \cite{SafeerOntoso2019}. Non-magnetic electrodes (Ti/Au) are also deposited both at the ends of the graphene channel and the MoTe$_2$ flake (labeled as 1-4 in Fig. 1a). Two-terminal resistance measurements as a function of temperature (Fig. 1d) confirm the semimetallic behavior of the MoTe$_2$ flake, as expected for its thickness \cite{song2018few}. The lack of signatures of structural phase transitions suggests that the MoTe$_2$ flakes remain in the 1T' phase down to 10 K~\cite{stiehl2019,zhang2016Raman}. 

To explore SCC phenomena in graphene/MoTe$_2$ heterostructures, non-local resistance measurements were performed as sketched in Fig. 1b. A current \textit{I}$_\textrm{C} =$ 10 $\mu$A is applied between the ferromagnetic  electrode (FM) and graphene (Fig. 1a,b). The spin accumulation generated at the FM-graphene interface diffuses along the graphene channel, which lies along the $x$-direction and can be absorbed at the graphene/MoTe$_2$ region. The spin current \textit{\textbf{j}}$_{s}$ diffuses along the \textit{z}-direction from graphene to MoTe$_{2}$ and along the \textit{x}-direction to the proximitized graphene. There, if SCC occurs, a non-local voltage \textit{V}$_\textrm{NL}$ is recorded along the MoTe$_2$ flake (Fig. 1a,b). It is important to note that, in our experimental geometry (Fig. 1b), only the charge current along the $y$-axis direction is detected. $R_{\textrm{NL}}=V_{\textrm{NL}}/I_{\textrm{C}}$ is, thus, proportional to the SCC resulting into charge current \textit{\textbf{j}}$_{c}$ along the $y$-direction. The orientation of the spin polarization \textit{\textbf{s}} reaching the MoTe$_2$/graphene region can be controlled along the \textit{x}- and \textit{y}- directions by switching or pulling the magnetization orientation of the FM injector by an external magnetic field. Additionally, if the field is perpendicular to the spin polarization, spins will precess while diffusing along the graphene channel. Thus, by changing the value of the applied magnetic field along $x$-, $y$- or $z$-directions we can study SCC for the three different spin polarization orientations ($s^x$,$s^y$ and $s^z$) at the graphene/MoTe$_2$ heterostructure. %(\textit{x}, \textit{y}, and \textit{z}). 

If the magnetic field is applied along the FM in-plane easy axis ($y$-direction), a spin current with $s^y$ will diffuse on the graphene channel without precessing. As $B_y$ increases, the magnetization of the FM is reversed and a square hystheresis loop appears in Fig. 2a, as the sign of \textit{R}$_{\textrm{NL}}$ switches at the coercive field of the FM. This unconventional signal -corresponding to the SCC of $s^y$ spins - has been already reported at room temperature in Ref. \cite{SafeerOntoso2019}. Conversion of spin polarization into a parallel charge current is forbidden by crystal symmetries both in MoTe$_2$ (see Supplementary Material) and in graphene. In Ref.\cite{SafeerOntoso2019}, this effect was attributed to symmetry breaking due to strain induced either in the fabrication process or at the van der Waals interface. This signal remains visible down to 10 K, but decreases in amplitude with temperature (Fig. 2a).

\begin{figure*}
\includegraphics[width=0.97\textwidth]{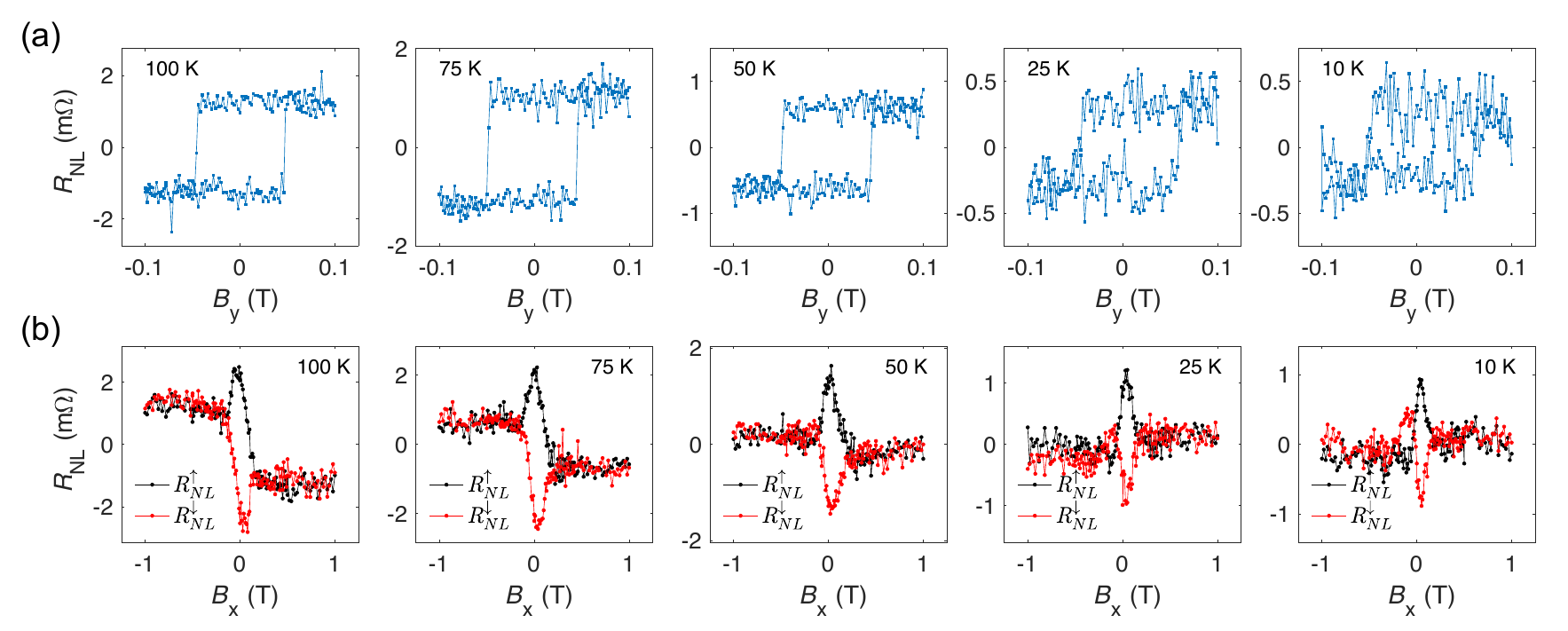}
\caption{\label{fig:2} Non-local resistance $R_{\textrm{NL}}$ measured with the configuration shown in Fig. 1b, while sweeping the magnetic field in (a) the in-plane FM easy axis $y$-direction, and (b) in the in-plane FM hard axis $x$-direction. In (b), curves are presented for initial magnetization of the Co electrode saturated along positive (black) and negative (red) $y$-direction. Measurements were performed at the temperature indicated in each panel.}
\end{figure*}

\begin{figure*}
\includegraphics[width=0.97\textwidth]{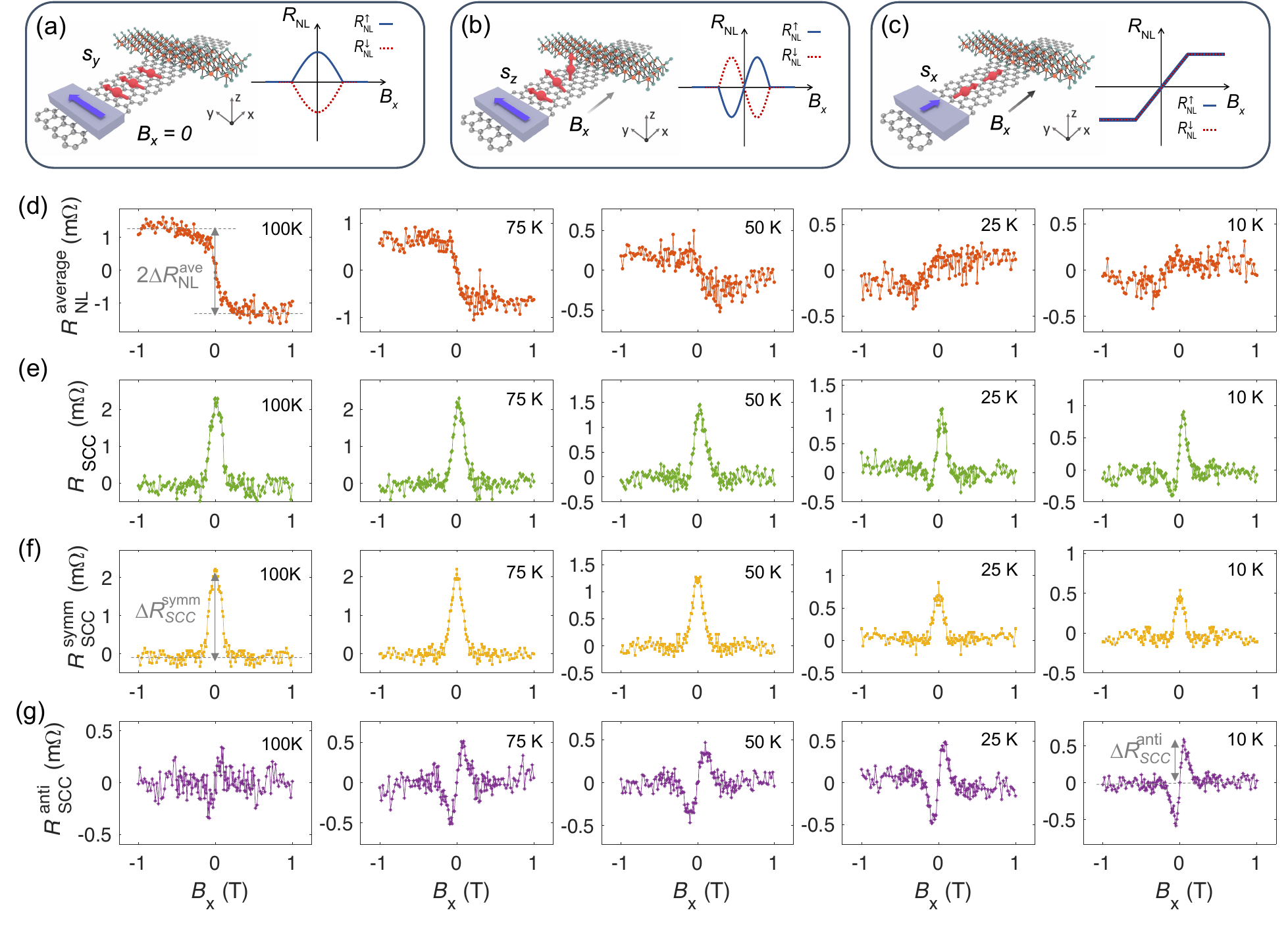}
\caption{\label{fig:3} (a-c) Sketches of the \textit{R}$_\textrm{NL}$ signal shapes associated to SCC from $s^y$ (a), $s^z$ (b) and $s^x$ (c). (d-g) Analysis of the \textit{R}$_\textrm{NL}$ versus $B_x$ curves in Fig. 2b at different temperatures:
(d) Average non-local resistance, defined as \textit{R}$_{NL}^{average} =$ (\textit{R}$_{NL}^{\uparrow} +$ \textit{R}$_{NL}^{\downarrow}$)/2. (e) Spin-to-charge conversion resistance defined as \textit{R}$_{SCC} =$ (\textit{R}$_{NL}^{\uparrow} -$ \textit{R}$_{NL}^{\downarrow}$)/2. (f) Symmetric component of the \textit{R}$_{SCC}$ signal obtained from data in panel e. (g) Anti-symmetric component of the \textit{R}$_{SCC}$ signal obtained from data in panel e. All signals are measured at the indicated temperatures.}
\end{figure*}

Further SCC components can be detected by sweeping the magnetic field in the in-plane hard axis of the FM injector, that is, along the $x$-direction ($B_x$). In Fig. 2b, %In this case, in a single experiment, the contributions to \textit{R}$_\textrm{NL}$ from conversion of the three spin polarization orientations ($s^x$, $s^y$, $s^z$) can be detected. 
\textit{R}$_\textrm{NL}$ is measured as a function of \textit{B}$_{x}$ for two different initial states of the FM magnetization. Prior to sweeping \textit{B}$_{x}$, the FM magnetization is first set along the $+$\textit{y}-direction [\textit{R}$_\textrm{NL}^{\uparrow}$, black curve in Fig. \ref{fig:2}b] %prior to each sweep of the field from 0 to 1 T and from 0 to $-$1 T. Lastly, the protocol is repeated but, in this case, the magnetization of the Co is initially set 
and then along the $-$\textit{y}-direction [\textit{R}$_\textrm{NL}^{\downarrow}$, red curve in Fig. \ref{fig:2}b]. This process is repeated for temperatures from 100 to 10 K. In Ref. \cite{SafeerOntoso2019}, such \textit{R}$_\textrm{NL}$ measurements were performed from 300 to 100 K (the data at 100 K is the same than in this publication) and the curves were interpreted as the sum of two contributions, one symmetric and another antisymmetric, corresponding, respectively, to the SCC of $s^x$ and $s^y$. This analysis was sufficient to provide an interpretation for the high temperature data. However, Fig. 2b shows how, at very low temperature, the \textit{R}$_\textrm{NL}$ signal not only changes in size, but it also undergoes significant changes in shape. This suggests the emergence of extra SCC contributions to \textit{R}$_\textrm{NL}$ that require of a more detailed analysis. To separate the SCC contributions from each spin orientation, it is crucial to consider how they depend on the initial state of the FM magnetization and their parity with respect to the applied magnetic field.
Initially, at zero magnetic field, only %injected 
spins polarized along the FM easy axis (\textit{s}$^{y}$) %(\textit{y}-direction)
will reach the graphene/MoTe$_2$ region and contribute to \textit{R}$_{\textrm{NL}}$ % The non-local resistance will
, with a sign that depends on the initial state of the FM magnetization, either along the $+$\textit{y} or the $-$\textit{y} direction (Fig. \ref{fig:3}a). As \textit{B}$_{x}$ increases, the amplitude of the spin \textit{s}$^{y}$ component decays due to spin precession in the \textit{y}$-$\textit{z}-plane (Fig. \ref{fig:3}b) and magnetization pulling of the injector along the \textit{x}-direction (Fig. 3c). As a consequence, the contribution to the \textit{R}$_{\textrm{NL}}$ signal coming from SCC of \textit{s}$^{y}$ is symmetric with respect to \textit{B}$_{x}$ and maximum at zero field (Fig. \ref{fig:3}a). Also, due to precession in the \textit{y}$-$\textit{z}-plane field, SCC of \textit{s}$^{z}$ at graphene/MoTe$_2$ can also be detected in the $B_x$ sweeps, with its contribution to \textit{R}$_{\textrm{NL}}$ being described by an antisymmetric Hanle curve %with a maximum at a certain value of \textit{B}$_{x}$ 
(Fig. \ref{fig:3}b).  This signal reverses for the two opposite initial states of the injector magnetization (Fig. \ref{fig:3}b). Finally, for large values of \textit{B}$_{x}$, the FM magnetization is aligned along the hard $x$-axis and \textit{s}$^{x}$ spins are injected, which do not precess when diffusing along the graphene channel (Fig. \ref{fig:3}c). \textit{R}$_{\textrm{NL}}$ due to SCC of \textit{s}$^{x}$ is linear at low fields, reaches saturation values at large magnetic field with opposite signs at $+$\textit{B}$_{x}$ and $-$\textit{B}$_{x}$ (S-shaped curve), and does not depend on the initial state of the injector magnetization  (Fig. \ref{fig:3}c). %SCC from $s^x$ is only dominant at large magnetic fields, when the magnetization of the Co is completely pulled in the direction of the B$_x$ field [see Fig. \ref{fig:3}c]. 

%On the one hand, the contribution of 
Because SCC from \textit{s}$^{x}$ is the only contribution to \textit{R}$_{\textrm{NL}}$ that does not depend on the initial state of the FM magnetization, it can be extracted by %removing any signal that depends on the initial state of the magnetization. That can be done by 
taking the average %of \textit{R}$_{NL}^{\uparrow}$ and \textit{R}$_{NL}^{\downarrow}$ as 
\textit{R}$_{NL}^{average} =$ (\textit{R}$_{NL}^{\uparrow} +$ \textit{R}$_{NL}^{\downarrow}$)/2 (Fig. 3d). The saturation values of \textit{R}$_\textrm{NL}^{average}$ as a function of $B_x$ change sign below 50 K. %signals measured at different temperatures display two saturating states above the saturation field of Co [Fig. \ref{fig:3}d], which %R$_\textrm{NL}^{average}$ 
%change sign below 50 K.
% comentado porque era un poco repetitivo con lo de abajo.  This signal is associated to SCC of \textit{s}$_{x}$ spins, as it was proven at higher temperatures via precession with \textit{B}$_{z}$ in Ref. \cite{SafeerOntoso2019} ({\color{red} añadir algo de esto a supp info? Discutir origen aqui? Podria ser conventional SHE con jz, ver apendice}. 
On the other hand, %the contribution to \textit{R}$_\textrm{NL}$ from SCC of \textit{s}$^{y}$ clearly depends on the initial state of the Co magnetization and has opposite sign for $+$\textit{s}$^{y}$ or $-$\textit{s}$^{y}$ spins [see Fig. \ref{fig:3}a]. Moreover, the contribution of \textit{s}$^{z}$ to \textit{R}$_\textrm{NL}$ will also depend on the initial state of the magnetization since it comes from $y$-spin precession. Therefore,
both SCC contributions from $s^y$ and $s^z$ are expected to change sign for opposite initial states of the FM magnetization. Therefore, these contributions are included in \textit{R}$_\textrm{SCC} =$ (\textit{R}$_\textrm{NL}^{\uparrow} -$ \textit{R}$_\textrm{NL}^{\downarrow}$)/2 (Fig. 3e).
While the \textit{s}$^{y}$ contribution is symmetric with respect to \textit{B}$_{x}$, the contribution of \textit{s}$^{z}$ is antisymmetric.~% with \textit{B}$_{x}$. 
 This difference allows us to establish an univocal relation between the SCC signal arising from \textit{s}$^{y}$, \textit{R}$_\textrm{SCC}^{symm}$ (Fig. 3f) and the SCC signal from \textit{s}$^{z}$, \textit{R}$_\textrm{SCC}^{anti}$ (Fig. 3g). \textit{R}$_\textrm{SCC}^\textrm{symm}$ shows a maximum at zero field and dissappears at the saturation value of \textit{B}$_{x}$ field [Fig. \ref{fig:3}f]. This SCC signal is the same one detected in the $B_y$ sweeps in Fig. 2a and is associated to SCC of \textit{s}$^{y}$ spins %, as it was proven at higher temperatures via precession with \textit{B}$_{z}$ 
in Ref. \cite{SafeerOntoso2019}.
%\textbf{revisar aqui es donde hay que cambiar un poco discusion}
More interestingly, at 10 K, \textit{R}$_\textrm{SCC}^\textrm{anti}$ displays a sharp antisymmetric Hanle curve. This signal progressively  disappears at higher temperatures, being still visible at 75 K [Fig. \ref{fig:3}g] and becoming nearly negligible at 100 K and above (see Supplementary Material and Ref. \cite{SafeerOntoso2019}). These signals evidence the SCC of \textit{s}$^{z}$ spins, which was not observed at higher temperature \cite{SafeerOntoso2019}. 

It is important to note that the signals in Fig. \ref{fig:3} are measured as a function of
\textit{B}$_{x}$ and, therefore, only \textit{R}$_\textrm{SCC}^\textrm{anti}$ (the \textit{s}$^{z}$ contribution) arises purely from spin precession, confirming its spin origin. In contrast, \textit{R}$_{SCC}^{symm}$ and 
\textit{R}$_{NL}^{average}$ are determined not only by precession, but partially or entirely by the pulling of the FM magnetization. %The SCC signal from \textit{s}$^{x}$ arises also from the pulling of the Co magnetization and no precession occurs.
%In order to prove the spin origin of the \textit{s}$^{x}$ and \textit{s}$^{y}$ SCC signals, spin precession measurements should be done by applying out-of-plane fields (\textit{B}$_{z}$). 
Spin precession measurements with applyed out-of-plane field (\textit{B}$_{z}$) were successfully performed at higher temperatures in Ref. \cite{SafeerOntoso2019}, confirming the spin origin of the SCC signals from \textit{s}$^{x}$ and \textit{s}$^{y}$, and discarding spurious effects \cite{safeer2021reliability}. However, at low temperatures, large backgrounds coming from the magnetoresistance of MoTe$_{2}$ and the Hall effect in graphene prevented us from identifying clear  Hanle signals. Nevertheless, since the low temperature evolution of the amplitudes of the SCC signals for \textit{s}$^{x}$ and \textit{s}$^{y}$ matches the trend at higher temperatures extracted from Ref.\cite{SafeerOntoso2019}) (see Supplementary Material), the spin origin of the \textit{R}$_\textrm{NL}^{average}$ and \textit{R}$_\textrm{SCC}^{symm}$ signals can be inferred. %Most importantly, the SCC signal from \textit{s}$^{z}$ arises from spin precession and, consequently, its spin origin is confirmed.

\begin{figure*}
\includegraphics[width=\textwidth]{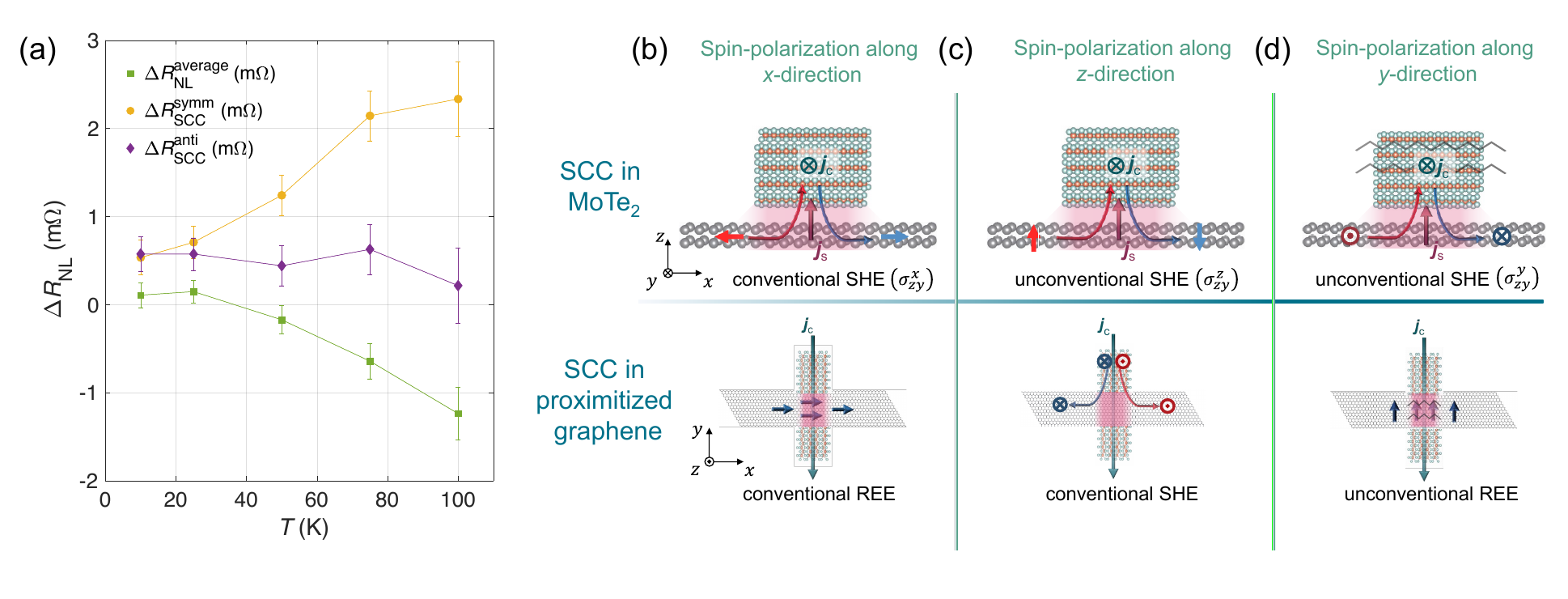}
\caption{\label{fig:4} a) Low-temperature dependence of the amplitude of %the three contributions to the \textit{R}$_\textrm{NL}$ signal, extracted in Fig. \ref{fig:3}. %The high temperature data, represented by squares, are extracted from Fig. \ref{temperature dependence amplitude}. 
\textit{R}$_\textrm{NL}^{average}$ %is associated to the 
(SCC of \textit{s}$^{x}$), 
\textit{R}$_{SCC}^{anti}$ %is the contribution arising from 
(SCC of \textit{s}$^{z}$), and \textit{R}$_\textrm{SCC}^{symm}$ (from SCC of \textit{s}$^{y}$) extracted from Fig \ref{fig:3}. Error bars are calculated from the standard deviation of the signal noise.% The inset is a zoom at low temperature. The dashed lined is a guide for the eyes at zero amplitude.  . 
(b-d) Summary of SCC processes in a graphene/MoTe$_{2}$ van der Waals heterostructure that can be detected in our experimental geometry (charge current along the $y$-direction). SCC can take place either at the bulk of the MoTe$_{2}$ or at the proximitized graphene region, which is represented by a pink halo. %When the charge current is along the \textit{y}-direction, all three spin polarization orientations can be generated. %Conversely, owing to Onsager reciprocity, all three spin polarization orientations would contribute to a charge current in the $y$-direction. %For simplicity, we discuss charge-to-spin conversion processes, instead of the reciprocal spin-to-charge conversion that has been experimentally detected. Spin polarization along the \textit{x}-direction can steam from conventional SHE in MoTe$_{2}$, 
(b) SCC of $s^x$ can occur via the conventional SHE in MoTe$_{2}$ %where \textit{\textbf{j}}$_{c} \perp$ \textit{\textbf{j}}$_{s} \perp$ \textit{\textbf{s}}, 
if a spin current is injected in the $z$-direction ($\sigma^x_{zy}$)% Alternatively, SCC of $s^x$ spins may occur 
or owing to a conventional REE in the proximitized graphene. % where \textit{\textbf{j}}$_{c} \perp$ \textit{\textbf{s}}. 
(c) SCC of %spin polarization along \textit{z}-direction 
$s^z$ can be attributed to an unconventional, non-orthogonal SHE in MoTe$_{2}$ ($\sigma^z_{zy}$) allowed by crystal symmetry %, where \textit{\textbf{j}}$_{c} \perp$ \textit{\textbf{j}}$_{s} ||$ \textit{\textbf{s}}. %This SHE component is allowed by the crystal symmetry of the 1T' phase 
%(see Supplementary Material) %. Nonetheless, $s^z$ conversion 
but can be also mediated by a conventional SHE in proximitized graphene% with \textit{\textbf{j}}$_{c} \perp$ \textit{\textbf{j}}$_{s} \perp$ \textit{\textbf{s}}. 
(d) SCC of $s^y$ (parallel to charge current), is allowed only if the mirror symmetry of 1T'-MoTe$_{2}$ or the ones of graphene/MoTe$_{2}$ interface are broken (represented by a zigzag line), % In that case, \textit{s}$^{y}$ spins can be converted either 
via unconventional, symmetry forbidden, SHE in MoTe$_{2}$ ($\sigma^y_{zy}$) %, where \textit{\textbf{j}}$_{c} ||$ \textit{\textbf{j}}$_{s} \perp$ \textit{\textbf{s}}, 
or via an unconventional EE in proximitized graphene. %, where \textit{\textbf{j}}$_{c} ||$ \textit{\textbf{s}}.
}
\end{figure*}

Shunting due to the presence of the MoTe$_2$ flake prevents us from determining the transport parameters for the proximitized graphene when fitting the data to a Hanle spin precession model. Still, we can extract the amplitude
for each SCC component as a function of temperature in the range from 10 K to 100 K, which are presented in  Fig. \ref{fig:4}a (see Supplementary Material for a comparison with the data at higher temperature from Ref. \cite{SafeerOntoso2019}). %The solid circles correspond to the amplitude at lower temperatures extracted from Fig. \ref{fig:3}d,e,f, and the open squares correspond to the amplitudes at higher temperatures, extracted from Safeer et al. \cite{SafeerOntoso2019}. If we focus only on the results at low temperature (inset in Fig. \ref{fig:4}), 
%The red circles represent 
The amplitude of \textit{R}$_\textrm{NL}^{average}$ (coming from SCC of $s^x$) changes sign at low temperature; being negative above 50 K and positive at 25 K and 10 K. At higher temperatures, in Ref. \cite{SafeerOntoso2019}, this contribution was considered to arise from conventional SHE in the bulk of MoTe$_{2}$, with $s^x$ spins been injected in the $z$-direction and resulting in a charge current along $y$. In other words, \textit{R}$_\textrm{NL}^{average}$ may correspond to a conventional configuration of the SHE where
where the electric field $E$, spin current $j_s$, and spin polarization $s^\alpha$ are mutually orthogonal ($j_{i}^{\alpha} = \sigma_{ij}^{\alpha} E_{j}$, with $\sigma_{zy}^{x}$, see Supplementary Material). However, the conventional REE at the proximitized graphene may also convey SCC in the $y$-direction from \textit{s}$^{x}$ (Fig. \ref{fig:4}b). Indeed, at low temperatures, the efficiency of proximity-induced REE in graphene, which is expected to be large \cite{ghiasi2019}, could overcome the bulk SHE in MoTe$_{2}$ and result in a change of sign of the overall detected signal. Therefore, the sign change of the $s^x$ signal with lowering T suggests a competition of effects, namely conventional SHE in MoTe$_2$ and conventional REE in proximitized graphene, with opposite sign in their efficiencies. Nevertheless, since we could not prove the spin origin of the low temperature component, we cannot discard that it might arise from artifacts similar to those observed in Ref. \cite{safeer2021reliability}.

Likewise, $\Delta$\textit{R}$_\textrm{NL}^{symm}$ (from SCC of $s^y$) decreases with lowering the temperature (Fig.\ref{fig:4}a). This can be also seen in Fig. \ref{fig:2}a where the amplitude of the \textit{R}$_\textrm{NL}$ hysteresis loop with \textit{B}$_y$ decreases with temperature and becomes barely visible at 10 K. Conversion of $s^y$ spins into a current in the parallel $y$-direction requires the reduction of the system symmetries. This signal can be attributed to an unconventional SHE in MoTe$_{2}$ with $\sigma_{zy}^{y}$, only if all mirror symmetries of MoTe$_{2}$ are broken, or to an unconventional REE in the proximitized graphene (Fig. \ref{fig:4}d). The hypothesis of a REE origin has recently gained force since the shear strain needed to break the MoTe$_{2}$ crystal symmetries is expected to be sample-dependent \cite{couto2014}, and this signal has been already observed in several different graphene/MoTe$_{2}$ heterostructures \cite{Ontoso_MoTe2_2022, SafeerOntoso2019}. Moreover, the misalignment between the mirror planes of graphene and MoTe$_{2}$ creates a non-symmetric interface, which enables the unconventional REE mechanism in proximitized graphene \cite{david2019,li2019_twist,naimer2021twist,peterfalvi2022quantum,veneri2022twist,lee2022charge}. 

%This scenario is represented in Fig. \ref{fig:4}b, where the zigzag lines in the proximitized area represent a non-symmetric interface. 

Interestingly, $\Delta$\textit{R}$_\textrm{NL}^{anti}$, observed only at low temperatures and coming from SCC of \textit{s}$^{z}$, presents its largest value at 10 K and decreases with increasing temperature. Two possible origins can lie behind the conversion of out-of-plane spins: unconventional SHE in MoTe$_{2}$, with \textit{\textbf{j}}$_{s} ||$ \textit{\textbf{s}} $\perp$ \textit{\textbf{j}}$_{c}$ and $\sigma_{zy}^{z}$ (see Supplementary Material), and conventional SHE in proximitized graphene (with \textit{\textbf{j}}$_{s} \perp$ \textit{\textbf{s}} $\perp$ \textit{\textbf{j}}$_{c}$) (Fig. \ref{fig:4}c). The non-orthogonal SHE in 1T'-MoTe$_{2}$ with spin polarization parallel to the spin current $\sigma_{zy}^{z}$, is allowed by the reduced crystal symmetry and has been reported in Ref. \cite{stiehl2019} with an efficiency $\sim$ 8 times smaller than the conventional SCC of $s^x$ ($\sigma^x_{zy}$, $\Delta$\textit{R}$_\textrm{NL}^{average}$ in this work). 
In contrast, SHE in proximitized graphene has been demonstrated to increase significantly with decreasing $T$ \cite{safeer2019,benitez2020}. Therefore, even if a non-orthogonal SHE in MoTe$_{2}$ cannot be fully discarded,
the rising of $\Delta$\textit{R}$_\textrm{NL}^{anti}$ at low temperatures with amplitude larger than $\Delta$\textit{R}$_\textrm{NL}^{average}$ points to a conventional SHE in proximitized graphene as the most likely origin for the SCC of \textit{s}$^{z}$ spins. %Nonetheless, a non-orthogonal SHE in MoTe$_{2}$, allowed by the reduced crystal symmetry of 1T'-MoTe$_{2}$, cannot be fully discarded as potential origin. 
Both the temperature at which the SCC signal from $s^z$ becomes visible and the temperature at which the SCC signal from $s^x$ changes sign seem to slightly vary from sample to sample (see Supplementary Material). These facts are compatible with their possible origin from conversion occurring in the proximitized graphene.

%Because SHE in proximitized graphene decreases with increasing $T$ \cite{safeer2019,benitez2020}, SCC from \textit{s}$^{z}$ spins is likely due to proximity-induced SHE in graphene. 
In conclusion, we observe omnidirectional SCC in graphene/MoTe$_{2}$ van der Waals heterostructures at low temperatures. Firstly, SCC from out-of-plane  \textit{s}$^{z}$ spins emerges at low temperatures, likely owing to proximity-induced SHE in graphene. This unconventional component, which is absent at room temperature, is allowed by the symmetries of the system. The SCC signal from \textit{s}$^{x}$, which reverses sign with \textit{T}, is compatible with a competition between mutually orthogonal SHE in MoTe$_{2}$ and REE in proximitized graphene, having opposite signs and varying differently in efficiency with temperature. The unconventional SCC from \textit{s}$^{y}$ spins requires mirror symmetries to be broken either at the MoTe$_{2}$, if mediated by unconventional SHE at the bulk of this material, or at the interface between graphene and MoTe$_{2}$, if the origin is unconventional REE in the proximitized graphene. %All the possibilities are summarized in Figure \ref{fig:4}b. 
Overall, the results presented in this work, together with those in Refs. \cite{SafeerOntoso2019, Ontoso_MoTe2_2022}, make graphene/MoTe$_{2}$ a fascinating heterostructure where a plethora of SCC processes arises from different mechanisms. In particular, the crystal structure of MoTe$_2$ makes this a unique system to further investigate the unconventional components enabled by both the low crystal symmetry of the TMD and those appearing in the proximitized graphene interface.%Although our results unveil a rich variety of possibilities, further studies in order to disentangle the exact origin of each SCC component will be required.

\section*{Supplementary Material}

Supplementary material contains Spin Hall conductivity tensor for 1T'-MoTe$_2$, a comparative of the amplitudes of SCC signals at higher temperature, Atomic Force Microscopy characterization of devices and SCC data from a second device.

\begin{acknowledgments}
This work is supported by the Spanish MICINN under Projects No. PID2021-122511OB-I00,
No. MAT2017-88377-C2-2-R, and the Maria de Maeztu Units of Excellence Programme (Grants No. MDM-2016-0618 and No. CEX2020-001038-M); the “Valleytronics” Intel Science Technology Center; the Gipuzkoa Regional Council under Projects No. 2021-CIEN-000037-01; and the European Union H2020 under the Marie Sklodowska-Curie Actions (Grants No. 0766025-QuESTech and No. 794982-2DSTOP). N.O.thanks the Spanish MICINN for support from a Ph.D.fellowship (Grant No. BES-2017-07963). J.I.-A. acknowledges support from the “Juan de la Cierva-Formación” program by the Spanish MICINN (Grant No. FJC2018-038688-I) for a postdoctoral fellowship. R.C. acknowledges funding from Generalitat Valenciana through Grants No. CIDEGENT/2018/004, IDIFEDER/2020/005 and IDIFEDER/2021/016.
\end{acknowledgments}

\section*{Data Availability Statement}
The data that support the findings of
this study are available from the
corresponding author upon reasonable
request.

\bibliography{Bibliography.bib}% Produces the bibliography via BibTeX.

\end{document}